\documentclass[twocolumn,prb,showpacs,preprintnumbers,amssymb,bibnotes,nofootinbib,runinaddress]{revtex4}
\usepackage{graphicx,epsfig,amsmath,amssymb}
\usepackage [latin1] {input enc}

\begin{document}

\title{Phase transitions and molecular dynamics of n-hexadecanol\\ confined in
silicon nanochannels}

\author{R.~Berwanger$^1$, A.~Henschel$^2$, K.~Knorr$^2$, P.~Huber$^2$, and R.~Pelster$^1$ \\
Universit{\"a}t des Saarlandes, $^1$FR 7.2 Experimentalphysik \& $^2$FR 7.3 Technische Physik, \\
 66041 Saarbr{\"u}cken, Germany}

\begin{abstract}
We present a combined x-ray diffraction and infrared spectroscopy
study on the phase behavior  and molecular dynamics of
n-hexadecanol in its bulk state and confined in an array of
aligned nanochannels of 8~nm diameter in mesoporous silicon. Under
confinement the transition temperatures between the liquid, the
rotator R$_{II}$ and the crystalline C phase are lowered by
approximately 20~K. While bulk n-hexadecanol exhibits at low
temperatures a polycrystalline mixture of orthorhombic $\beta$-
and monoclinic $\gamma$-forms, geometrical confinement favors the
more simple $\beta$-form: only crystallites are formed, where the
chain axis are parallel to the layer normal. However, the
$\gamma$-form, in which the chain axis are tilted with respect to
the layer normal, is entirely suppressed. The $\beta$-crystallites
form bi-layers, that are not randomly orientated in the pores. The
molecules are arranged with their long axis perpendicular to the
long channel axis. With regard to the molecular dynamics, we were
able to show that confinement does not affect the inner-molecular
dynamics of the CH$_2$ scissor vibration and to evaluate the
inter-molecular force constants in the C phase.
\end{abstract}

 \maketitle

\section{Introduction}
The physical properties of condensed matter spatially confined in
pores or channels of a few  nanometer in diameter can differ
markedly from the behavior in the bulk state. In particular, phase
transitions can be entirely suppressed or significantly altered in
comparison to their bulk counterparts
\cite{Gelb1999,AlbaSim2006,Christenson2001,Knorr2008}. Also the
dynamics of condensed matter confined in mesopores, most
prominently in the vicinity of glass transitions
\cite{Koppensteiner2008,Scheidler2000,Kremer1999,Jackson1991,Barut98,Pelster99prb,Daoukaki98prb,Pissis98,
Schranz2007, Frick2003}, can be affected markedly.

Intimately related to these changes in the phase transition
phenomenology the architectural  principles of molecular solids
can substantially differ in the spatially confined state from the
bulk state. This depends, however, sensitively on the complexity
of the building blocks. For simple van-der-Waals systems, such as
Ar and N$_2$, a remarkable robustness of the bulk structures has
been found for the solid state under confinement \cite{Huber1998,
Wallacher2001, Knorr2003}. By contrast, the structural properties
of pore fillings built out of more complex building blocks, such
as linear hydrocarbons
\cite{Huber2006,Henschel2007,Montenegro2003,Xie2008,Valliulin2006}
or liquid crystals \cite{Crawford1996,Kityk2008} are very
susceptible to confinement on the meso- and nanoscale. For
example, a quenching of the lamellar ordering of molecular
crystals of n-alkanes has been observed in tortuous silica
mesopores of Vycor \cite{Huber2004}. However, in tubular channels
of mesoporous silicon this building principle of hydrocarbon
molecular crystals survives, albeit a peculiar texture has been
observed for the pore confined solids \cite{Henschel2007}: The
long axes of the molecules and thus the stacking direction of the
lamellae are oriented perpendicular to the long axis of the pores.

Here we present an experimental study on a medium-length, linear
alcohol C$_{16}$H$_{33}$OH,  a representative of the 1-alcohol
series, imbibed in mesoporous silicon. We explore the phase
behavior of the confined alcohol by a combination of x-ray
diffraction and infrared spectroscopy measurements. As we shall
demonstrate, we profit in those experiments both from the parallel
alignment of the silicon channels and from the transparency of the
silicon host in the infrared region.

\section{Experimental}
The porous silicon samples used in this study were prepared by
electrochemical etching  of a heavily p-doped (100) silicon wafer
\footnote{producer: SiMat, Landsberg, Germany; specific
conductivity: $\rho=0.01-0.025$~$\Omega$cm.} with a current
density of 13~$\frac{mA}{cm^2}$ in a solution composed of HF,
ethanol and H$_{2}$O (1:3:1 per volume) \cite{Lehmann1991,
Zhang2000, Cullis1997}. These conditions led to a parallel
arrangement of non-interconnected channels oriented with their
long axes along the $<$100$>$ crystallographic direction of
silicon, which coincides with the normal of the wafer surface.
After the porous layer had reached the desired thickness of
70~microns, the anodization current was increased by a factor of
ten with the result that the porous layer was released from the
bulk wafer underneath. Using nitrogen sorption isotherms at
$T=77$~K, we determined a porosity of 60\% and a mean channel
diameter of 8~nm. The single crystalline character of the matrix
was checked by x-ray diffraction. Transmission electron
micrographs of channel cross sections indicate polygonal, rough
channel perimeters rather than circular, smooth circumferences
\cite{Gruener2008}.

The matrix both for the infrared spectroscopy and the x-ray measurements
were filled completely via capillary action (spontaneous imbibition) with
liquefied C$_{16}$H$_{33}$OH \cite{Huber2007}. Bulk excess material at the surface was removed by paper tissues.\\

Infrared spectra in a range of wavenumbers $\overline{\nu}$ from
4000 to 800~cm$^{-1}$ with a resolution of 1~cm$^{-1}$ were
measured with a Fourier Transform Spectrometer (FTIR Perkin Elmer
System 2000). This range corresponds to frequencies from
$3\cdot10^{13}$~Hz to $1.2\cdot10^{14}$~Hz (wavelengths from
10~$\mu$m to 2.5~$\mu$m). For both the bulk material and the
filled porous samples the same sample holder was used, i.~e. a
copper cell with two transparent KBr windows. In the confinement
experiments the long channel axes were oriented parallel to the
beam axis, i.~e.\ perpendicular to the electric field vector. The
sample holder was placed into a cryostat (a closed cycle
refrigerator CTI cryogenics, Model 22) allowing us to vary the
temperature from 50 to 340~K. The temperature was controlled with
a LakeShore 340 temperature controller with a precision of $\pm
0.25$~K. All IR-spectra that we show in the following were
measured during cooling {\bf (typical cooling rates were of the
order of 0.5~K/min)}. Heating scans show the same
behavior except for the transition temperatures, which are some degrees higher (see below).\\

\begin{figure}[hbt]
\begin{center}
\includegraphics[scale=0.35]{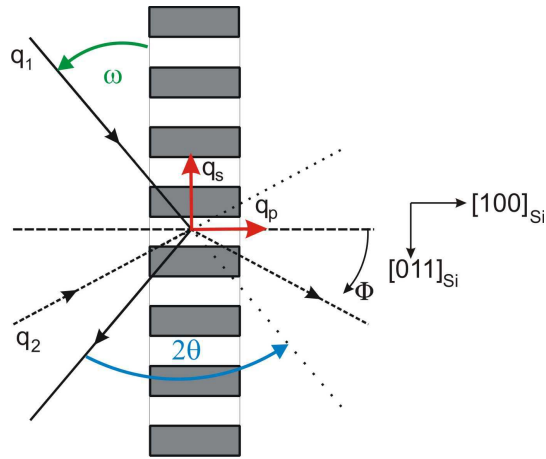}
\caption{\label{realRaum1} \small{(Color online) The angular variable $\Phi$
($\Phi = \omega - \theta$ with $\Phi$= 0° for scans along q$_{\rm p}$ and
$\Phi$ = 90° for scans along q$_{\rm s}$) is indicated. Also shown are the
incoming and outgoing x-ray beam for scans along q$_{\rm p}$ (solid lines:
wave vector component parallel to the pore axis, this means reflection geometry),
 and along q$_{\rm s}$ (dashed lines: wave vector component perpendicular to the
 pore axis; this is transmission geometry). For scans along q$_{\rm p}$, the
  detector angle $2\theta$ and the rotation angle $\omega$ are represented. }}
\end{center}
\end{figure}

For the x-ray measurements the sample was mounted on a frame in a sample cell consisting of a
Peltier cooled base plate and a Be cap. The cell was filled with He gas for better thermal contact.
The Be cap sits in a vacuum chamber, the outer jacket of which has Mylar windows allowing the passage
of the x-rays over a wide range of scattering angles $\theta$ within the scattering plane
(see Fig.~\ref{realRaum1}). But the set-up allowed practically no tilt with respect to
the scattering plane. The temperature was controlled by a LakeShore 330 over an accessible
range from 245~K up to 370~K. The measurements were carried out on a two-circle x-ray
diffractometer with graphite monochromatized CuK$_{\rm \alpha}$ radiation emanating
from a rotating anode. The porous sheet was mounted perpendicular to the scattering plane.
The two angles that could be varied were the detector angle $2\theta$ and the rotation angle
 $\omega$ about the normal of the scattering plane. The samples were studied as a
 function of temperature by performing several $\Phi$-scans. In this paper we
 concentrate on radial $2\theta$-$\omega$-Scans in reflection geometry, i.e.
 along q$_{\rm p}$ with $\Phi$=0°, and in transmission geometry, i.e. along
 q$_{\rm s}$ with $\Phi$=90° (see Fig.~\ref{realRaum1}).

\subsection{Structure of bulk n-hexadecanol}
\label{sec:bulkstruct}
 n-Hexadecanol, C$_{16}$H$_{33}$OH, is an
almost rod-like molecule with a length of 22~\r{A} and a width of
4~\r{A}. The C-atoms of the backbone are in an
all-trans-configuration so that they are located in a plane
\cite{Huber2004}.

\begin{figure}[hbt]
\begin{center}
\includegraphics[scale=0.36]{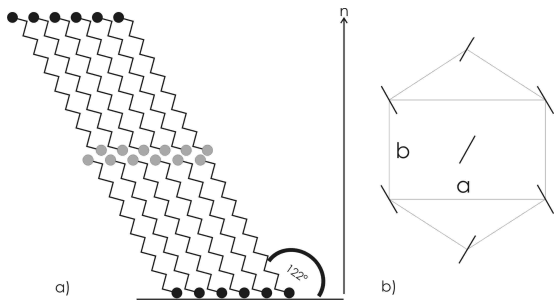}
\caption{\label{fig:bulkstructure_cryst} $\gamma$-form of the
crystalline low temperature phase of bulk C$_{16}$H$_{33}$OH ($T \le
310$~K). The structure is monoclinic. The left sketch shows the
orientation of the molecules with respect to the layer normal $n$,
the right sketch the in-plane arrangement, i.~e.\ a projection of
the backbones into the a-b-plane. Compare with the $\beta$-form
sketched in Fig.~\ref{fig:confstructure_cryst}}
\end{center}
\end{figure}

\begin{figure}[ht]
\begin{center}
\includegraphics[scale=0.3]{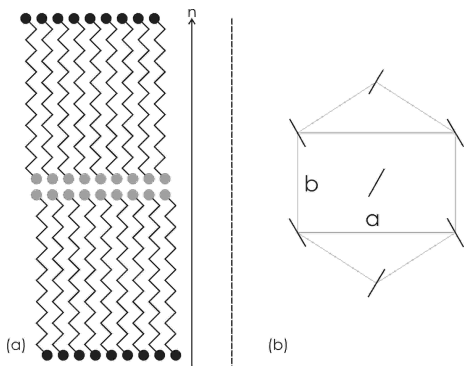}
\caption{\label{fig:confstructure_cryst} $\beta$-form of the
crystalline low temperature phase of C$_{16}$H$_{33}$OH. In contrast to
the $\gamma$-form (see Fig.~\ref{fig:bulkstructure_cryst}a), the
long chain axes are not tilted but parallel to the layer normal
$n$, i.~e.\ the structure is orthorhombic. Bulk
C$_{16}$H$_{33}$OH can exhibit a polycrystalline mixture of $\gamma$- and
$\beta$-form (see Sec.~\ref{sec:bulkstruct}). Confinement into
nanopores leads to the $\beta$-form (see below,
Sec.~\ref{sec:confstruct}).}
\end{center}
\end{figure}

At low temperatures n-alcohols form bi-layered crystals in two
possible  modifications: the so-called $\gamma$-form, i.~e.\ a
monoclinic structure as sketched in
Fig.~\ref{fig:bulkstructure_cryst} ($C_{2h}^{6}-A2/a$
\cite{Metivaud2005,Abrahamsson1960}), or the so-called
$\beta$-form, i.~e.\ an orthorhombic structure as sketched in
Fig.~\ref{fig:confstructure_cryst} \cite{Tasumi1964}. In the
$\gamma$-form, the molecules include an angle of $122$° with the
layer plane. Within the layers, they are close-packed in a
quasi-hexagonal 2D array, described by the rectangular in-plane
lattice parameters $a$ and $b$ (according to
Ref.~\cite{Abrahamsson1960} $a=7.42$~\r{A} and $b=4.93$~\r{A}
holds, so that $a/b=1.5$). There are two different alternating
orientations for the C-C-plane of the backbone leading to a
herringbone structure (see Fig.~\ref{fig:bulkstructure_cryst}b).
The $\beta$-form exhibits an identical orientational order of the
backbone, but the molecules' axes remain perpendicular to the
layers as sketched in Fig.~\ref{fig:confstructure_cryst}
\cite{Tasumi1964}. In addition, gauche- and trans-conformation of
the CO-bond alternate with molecules in this phase, while they are
in an all-trans configuration in the $\gamma$-form. In general,
the $\gamma$-form dominates at low temperatures for the even
alcohols, while the $\beta$-form is more frequent in odd
n-alcohols \cite{Ventola2002,Tasumi1964}. For n-hexadecanol both
the orthorhombic $\beta$-form \cite{Tasumi1964} and the monoclinic
$\gamma$-form \cite{Metivaud2005,Abrahamsson1960} are reported.
Depending on the preparation conditions it is possible to obtain a
polycristalline mixture of the monoclinic $\gamma$- and the
orthorhombic $\beta$-form \cite{Ventola2002}.

\begin{figure}[ht]
 \center\includegraphics[scale=0.3]{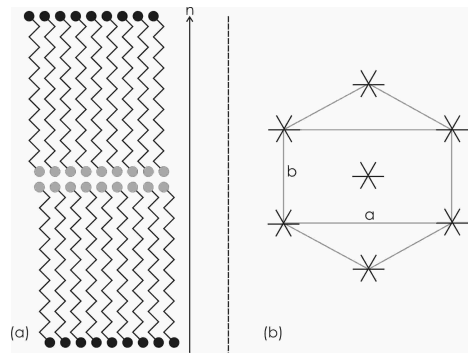}
 \caption{Structure of bulk n-hexadecanol in the Rotator-(II)-phase (for $310 \le T \le
 322$~K), a hexagonal arrangement.
The right picture shows the perfect hexagonal lattice in the
a-b-plane. Confined C$_{16}$H$_{33}$OH exhibits the same structure in its
rotator phase, but in a different temperature range (see below,
Table~\ref{tab:MeltingPoints}).
 \label{fig:bulkstructure_rot}}
\end{figure}

Upon heating, the crystalline phase undergoes a transition into a
so-called  Rotator-(II)-phase $R_{II}$, which is schematically
depicted in Fig.~\ref{fig:bulkstructure_rot} \footnote{For several
alkanes, there also exists a Rotator-(I)-Phase $R_I$, where the
molecules switch between two equal positions.}. This phase has a
hexagonal in-plane arrangement with the $c$-direction
perpendicular to the cell base. The hexagonal arrangement can be
indexed with an orthorhombic cell with a ratio of rectangular
basal lattice parameters of $a/b=\sqrt{3}$~ \cite{Sirota1996}. On
a microscopic level the change in the center of mass lattice from
the low-temperature crystalline phase to the rotator phase can be
attributed to the onset of hindered rotations of the molecules
about their long axes between six equivalent positions (the stars
in Fig.~\ref{fig:bulkstructure_rot}b). Further heating above 322~K
leads to the liquid state \cite{Sirota1996}.

\section{Results}

\subsection{Structure of confined n-hexadecanol}

\label{sec:confstruct}

\begin{figure}[ht]
\begin{center}
\includegraphics[scale=0.5, angle=0]{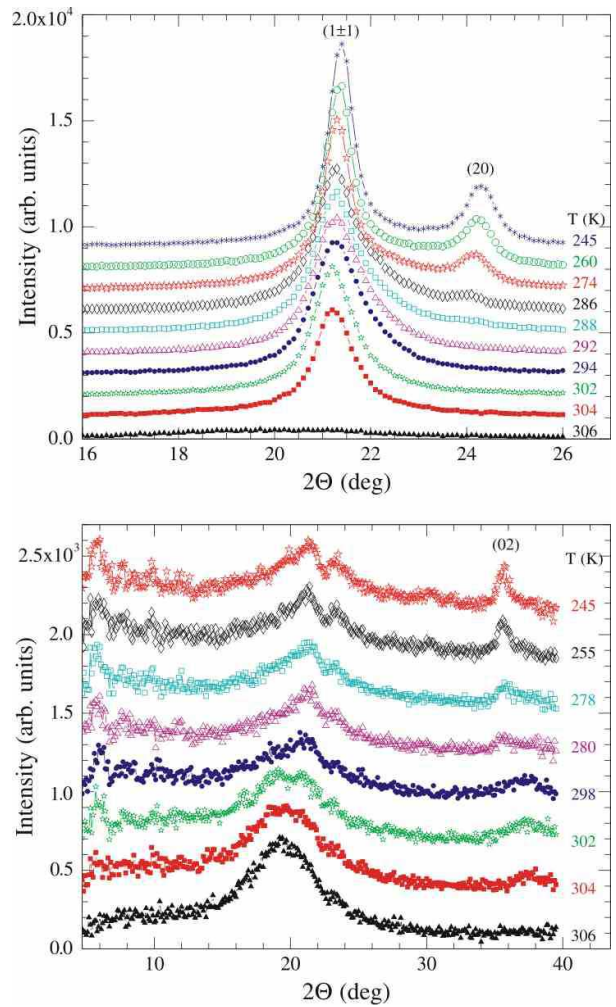}
\caption{\label{realRaum} \small{(Color online) Upper panel:
Diffraction pattern of q$_{\rm p}$-scans of pore confined
C$_{16}$H$_{33}$OH while cooling at selected temperatures. Lower
panel: Diffraction pattern of q$_{\rm s}$-scans at selected
temperatures while cooling. The leading in-plane reflections are
barely visible. Moreover, at lower angles ($2\theta\leq10$) there
are very weak and diffuse layering reflections.}}
\end{center}
\end{figure}

We have determined structures, phase sequences and transition
temperatures of  n-C$_{16}$H$_{33}$OH confined in mesoporous
silicon by x-ray diffractometry. The upper panel in
Fig.~\ref{realRaum} shows diffraction patterns along q$_{\rm p}$
at selected temperatures while cooling. The appearance of a broad
Bragg peak at $2 \theta \simeq 21$° indicates solidifaction. Its
position is compatible with the leading hexagonal in-plane
reflection of the $R_{II}$ phase. Upon further cooling a second
peak at $2 \theta \simeq 24$° shows up. This change in the
diffraction pattern indicates an uniaxial deformation of the
hexagonal lattice. Both reflections can be mapped on a 2D
rectangular mesh characteristic of an uniaxially deformed
hexagonal cell. The overall resulting pattern is, however,
incompatible with the monoclinic structure of the low temperature
bulk crystalline phase.

Additionally to the q$_{\rm p}$-scans, we performed also scans for
a variety of additional orientations of the scattering vector with
regard to the long axis of the channels. These patterns differ
markedly, which is indicative of a strong texture of the pore
confined cystallized alcohol. It is no powder in the
crystallographic sense. In particular, there are strong in-plane
reflections and no layering reflections for scans along q$_{\rm
p}$, while the q$_{\rm s}$-scans for the same sample show at least
very weak reflections characteristic of a bi-layer stacking and
only very weak leading in-plane reflections (see
Fig.~\ref{realRaum}). An analysis of the width of the layering
reflections yields a coherence length of 7\,($\pm1.5$)\,nm.

As discussed in more detail in Refs. \cite{Henschel2007}  and
\cite{Henschel2008}, the overall picture which emerges from these
results can be summed up as follows: the alcohol molecules form
orthorhombic structures with a bilayer-stacking direction along
the $c$-direction. Within the bilayers (the a-b-plane), the
molecules' backbones are untilted with regard to the stacking
direction and the backbones are orientationally either fully
ordered (in a herringbone fashion) or partially ordered, as known
from the R$_I$ phase of n-alkanes. The superlattice reflection
characteristic of the full, herringbone type orientational
ordering has been searched for and could weakly be detected at low
temperatures. The degree of uniaxial deformation of the hexagonal
center of mass cell, quantified by the deviation of the ratio
$a/b$ from its value in the hexagonal phase ($\sqrt{3}$), also
indicates a full orientational ordered state (see Table I,
\cite{Abrahamsson1960}). Thus, the diffraction data are compatible
with the bulk $\beta$ modification discussed above. This
conclusion is also supported by an analysis of the infrared
spectroscopy data sets presented below.

\begin{figure}[ht]
\begin{center}
\includegraphics[scale=0.35]{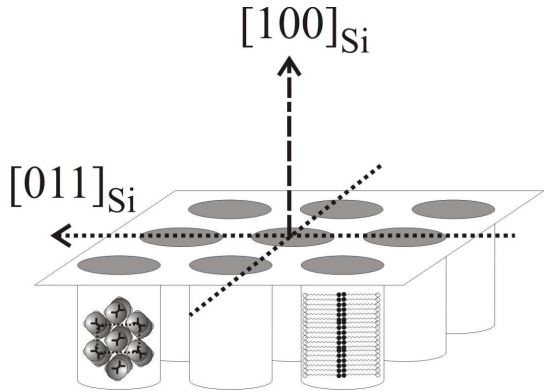}
\caption{\label{Porenschnitt} \small{Sketch of the orientation  of
pore condensed C$_{16}$H$_{33}$OH in the porous sheet. Left pore:
view on the rectangular in-plane arrangement of the molecules.
Right pore: bi-layered crystal; the long molecule axis are
oriented perpendicular to the pore axes. At maximum two bi-layers
fit into a pore.}}
\end{center}
\end{figure}

More importantly, the peculiar dependency of the diffraction
patterns on the orientation of the q-vector with regard to the
silicon host indicate that the bi-layer stacking direction is
perpendicular to the long axis of the channels and, consequently,
that the long axis of the molecules is oriented perpendicular to
the long axis of the channels (see Fig.~\ref{Porenschnitt}). At
first glance, this finding may appear somewhat counter-intuitive.
Albeit it can be understood as resulting from the crystallization
process in a strongly anistropic, capillary-like confined liquid
\cite{Henschel2007, Steinhart2006}. It is a well established
principle in single crystal growth that in narrow capillaries the
fastest growing crystallization direction prevails over other
directions and propagates along the long axes of capillaries
\cite{Palibin1933}. For layered molecular crystals of rod-like
building blocks this direction is an in-plane direction, which is
perpendicular to the long axis of the rods. If this direction is
aligned parallel to the silicon nanochannels due to the
crystallization process, it dictates a perpendicular arrangement
of the molecules' long axes with regard to the long channel axis,
in agreement with the diffraction results presented here.

\begin{table}[ht]
\begin{center}
\begin{tabular}{|c||c|c|c|}\hline
   & \multicolumn{1}{|c|}{bulk} & \multicolumn{2}{|c|}{confinement}\\\cline{2-4}
   & cryst. & R$_{II}$ & cryst.\\\hline \hline
 a [\AA] & 7.42 & 8.35 & 7.33\\\hline
 b [\AA] & 4.93 & 4.82 & 5.04\\\hline
 a/b & 1.51 & $\sqrt{3}$ & 1.45\\\hline
 d [\AA] & 8.91 & 9.64 & 8.90\\\hline
\end{tabular}
\end{center}
\caption{\label{tab:ab} Lattice parameters a and b of bulk and
confined C$_{16}$H$_{33}$OH and the diagonal $d=\sqrt{a^2+b^2}$ of the
subcell (see Fig.~\ref{fig:lattice}). The confined data result
from our x-ray measurements and the bulk data are taken from the literature \cite{Abrahamsson1960}.}
\end{table}

The temperature dependent diffraction study allows us to  gain
additional information on the relative stability of the different
nanochannel confined phases. In Table~\ref{tab:MeltingPoints} we
display the phase transition temperatures of confined
C$_{16}$H$_{33}$OH as inferred from the appearance or
disappaerance of characteristic Bragg peaks. There is a hysteresis
between heating and cooling for both the fluid-R$_{II}$- and the
R$_{II}$-C-transition (8~K and 3~K, respectively). Compared to the
bulk data (see also Tab.~\ref{tab:MeltingPoints}), the transition
temperatures of pore confined C$_{16}$H$_{33}$OH are lowered. On
cooling, the lowering is of the order of $\Delta T$= 18~K for the
fluid-$R_{II}$-transition and $\Delta T$= 26~K for the
$R_{II}$-C-transition. This observation is analogous to phase
transitions shifts reported for other pore
condensates~\cite{Christenson2001,AlbaSim2006}.

Furthermore, the temperature range of the confined $R_{II}$ phase,
14~K upon cooling and 19~K upon heating, is larger than that of
the bulk material (12~K). Obviously, confinement stabilizes the
orientational disordered $R_{II}$ phase, similarly as has been
found for n-alkanes~\cite{Henschel2007} and for other
orientational disordered, plastic phases under spatial confinement
\cite{Knorr2008}.

\begin{table}[ht]
\begin{center}
\begin{tabular}{|l||c|c|c|c|}
\hline
& fluid - R$_{II}$ & R$_{II}$ - C & fluid - R$_{II}$ & R$_{II}$ - C\\
\hline \hline
confined (cooling) & 304 & 291 & & \\
\hline
confined (heating) & 312 & 293 & & \\
\hline
bulk &  &  & 322 & 310\\
\hline
\end{tabular}
\caption{\label{tab:MeltingPoints} \small{Transition  temperatures
as determined by x-ray experiments of confined and bulk
C$_{16}$H$_{33}$OH \cite{Sirota1996}. They agree with those
inferred from IR - measurements (CH$_2$-scissoring vibration), see
Figs.~\ref{fig:CHspectrum} and \ref{fig:CHTemp}.}}
\end{center}
\end{table}

\begin{figure}[htb]
 \center\includegraphics[scale=0.35]{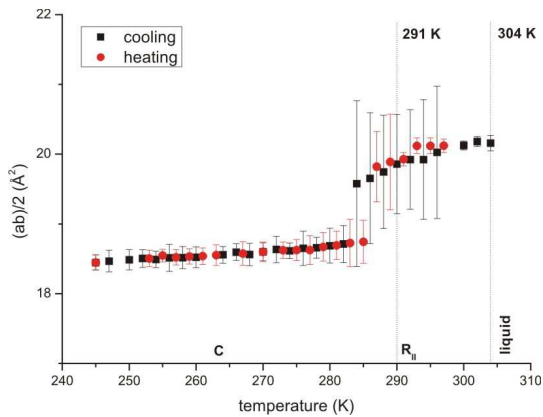}
 \caption{(color online). Temperature dependence of the area
 per molecule $A$ for C$_{16}$H$_{33}$OH
confined in porous silicon.}
 \label{fig:GitterparameterFlaeche}
\end{figure}

Since the pores were completely filled at higher temperatures,
when  hexadecanol is in its liquid state, the pore filling at low
temperatures does not consist only of bi-layer crystals: the
change of volume at the R$_{II}$-C phase transition is about 10\%
(see Fig.~\ref{fig:GitterparameterFlaeche}), so that there are
voids and/or molecules that are not part of a bi-layer crystal.
However, our experiments do not give us information about their
spatial arrangement.

\subsection{Molecular dynamics}

The dynamics of bulk-C$_{16}$H$_{33}$OH has already been
investigated in IR-measurements in the past
\cite{Metivaud2005,Tasumi1964}. In order to show later on how the
molecular dynamics is affected by spatial confinement on the
nm-scale, we display some of our bulk spectra in the following.
Here we focus on two characteristic vibrations,
the OH-stretching and the CH$_2$-scissoring vibration.\\

\begin{figure}[ht]
\begin{center}
\includegraphics[scale=0.43]{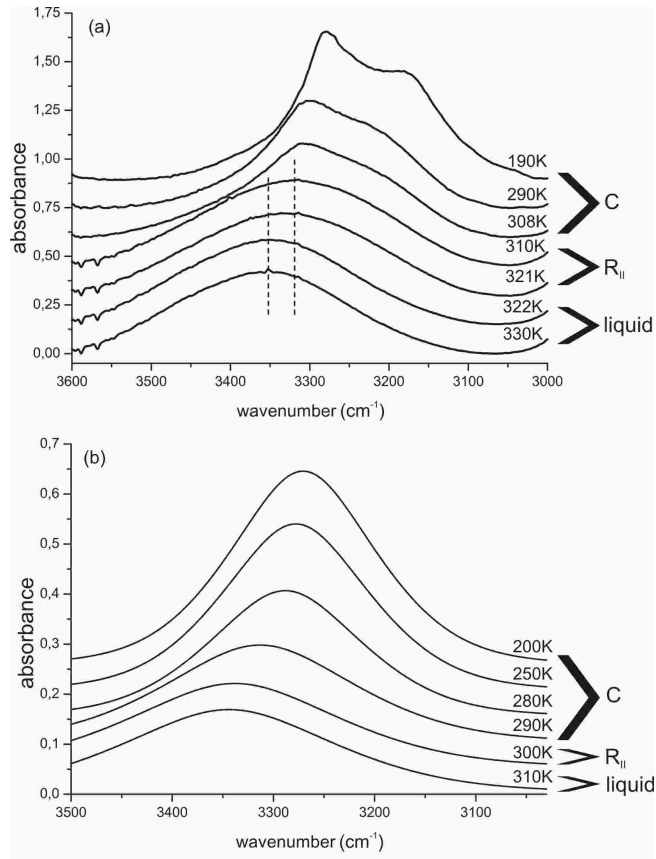}

 \caption{(a) IR spectrum in the OH - stretching range for bulk C$_{16}$H$_{33}$OH.
 At lower temperatures the peak shifts to lower wavenumbers and then splits into two
 peaks. (b) Spectrum for confined C$_{16}$H$_{33}$OH, where no splitting
 is visible.
\label{fig:OHspectrum}}
\end{center}
\end{figure}

\begin{figure}[ht]
\center\includegraphics[scale=0.45]{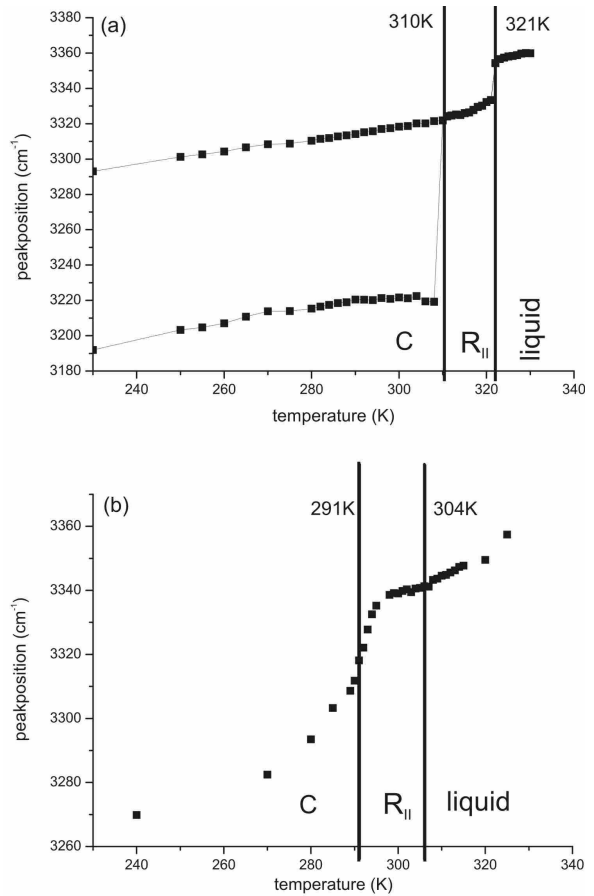} \caption{(a)
Wavenumber $\omega/2\pi c$ of the OH - stretching peak vs
temperature for bulk C$_{16}$H$_{33}$OH [compare with
Fig.~\ref{fig:OHspectrum}a)]. Three different phases are visible:
a) above 321~K, b) from 310 to 321~K, where the peak position
appears at lower wavenumbers and c) below 310~K where the peak
splits up into two peaks. (b) Wavenumber of the OH - stretching
peak for confined C$_{16}$H$_{33}$OH (compare with
Fig.~\ref{fig:OHspectrum}b). The transition between the
C and the R$_{II}$ phase seems to be smeared in a range
around $T=291 \pm 5$~K. The R$_{II}$-liquid transition does not
affect the OH-stretching.
 \label{fig:OHTemp}}
\end{figure}

Figs.~\ref{fig:OHspectrum}a) and \ref{fig:OHTemp}a) show  the bulk
spectra of the OH-stretching-band in the respective phases
(compare with
Figs.~\ref{fig:bulkstructure_cryst}-\ref{fig:bulkstructure_rot}).
In the liquid state (above 322~K) the peak maximum is located at
about 3345~cm$^{-1}$. A decrease of temperature below 321~K yields
a shift of the peak position to about 3325~cm$^{-1}$ indicating
the molecular rearrangement in the $R_{II}$ phase. A further
decrease of temperature below 310~K results in a splitting into
two peaks at approximately 3310~cm$^{-1}$ and 3220~cm$^{-1}$.
Confined C$_{16}$H$_{33}$OH shows a different behavior. There is
only one peak in the whole temperature range, the position of
which changes reflecting the transition between liquid phase and
$R_{II}$ phase as well as between $R_{II}$ phase and C phase (see
Figs.~\ref{fig:OHspectrum}b and \ref{fig:OHTemp}b).

The fact that the OH-band of bulk C$_{16}$H$_{33}$OH splits at low
temperatures while no splitting is observed under confinement
confirms the structural differences already observed in the x-ray
experiment. For example, Tasumi et. al have studied bulk alcohols
C$_n$H$_{2n+1}$OH from $n=11-37$ using infrared spectroscopy
\cite{Tasumi1964}, Ventòla et al. alcohols with $n=17-20$
\cite{Ventola2002}. Those alcohols showing at low temperatures (C
phase) the monoclinic $\gamma$-form, such as C$_{16}$H$_{33}$OH,
exhibit the splitting of the OH-band, while those that take the
orthorhombic $\beta$-form show a single peak. This is due to
differences in the spatial arrangement of the hydrogen bonds as
well as in the distances of neighboring O-atoms: in the
crystalline $\gamma$-form, where the molecule axis are tilted (see
Fig.~\ref{fig:bulkstructure_cryst}), the molecules show an all
trans conformation, and the intra-layer O-distance ($\simeq
2.74$~\r A) differs from the inter-layer O-distance ($\simeq
2.69$~\r A). However, in the orthorhombic $\beta$-form
(Fig.~\ref{fig:confstructure_cryst}) trans- and gauche-molecules
alternate and the intra-layer O-distance ($2.73$~\r A) nearly
equals the inter-layer O-distance ($2.72$~\r A), so that the
splitting is suppressed \cite{Tasumi1964}. Therefore, the observed
OH-band splitting for bulk C$_{16}$H$_{33}$OH shows the presence
of the $\gamma$-form. Either the whole bulk material exhibits the
$\gamma$-form or there is a mixture of $\gamma$- and
$\beta$-crystallites. The latter case is frequently
observed~\cite{Tasumi1964,Ventola2002}: in fact, in the range of
wavenumbers from 1150~cm$^{-1}$ to 950~cm$^{-1}$, where C-C
stretching vibrations are visible, we see indications for a
superposition of both forms (not shown). On the other hand, pore
confined C$_{16}$H$_{33}$OH shows no OH-band-splitting at low
temperatures. This reflects that the molecular arrangement doesn't
transform in the monoclinic $\gamma$-form but remains in an
orthorhombic structure, i.~e.\ only the $\beta$-form is present
(compare Figs.~\ref{fig:bulkstructure_cryst} and
\ref{fig:confstructure_cryst}). This result is in agreement with
the x-ray data presented above. {\bf Upon cooling, both the bulk
and the confined hexadecanol pass from an hexagonal $R_{II}$-
phase into a crystalline phase. The bulk material undergoes a
stronger structural change, i.~e.\ there is a mixture of the
orthorombic $\beta$- and  the monoclinic $\gamma$-form. The latter
one is suppressed under confinement, so that only the $\beta$-form
remains, which is quite similar to the hexagonal structure of the
R$_{II}$-phase:} the fact that the crystallites have to fit into
nanopores of irregular shape might favor the geometrically more
simple $\beta$-form \cite{Christenson2001,Morishige2000}(see
Fig.~\ref{Porenschnitt}).

\begin{figure}[ht]
\center\includegraphics[scale=0.43]{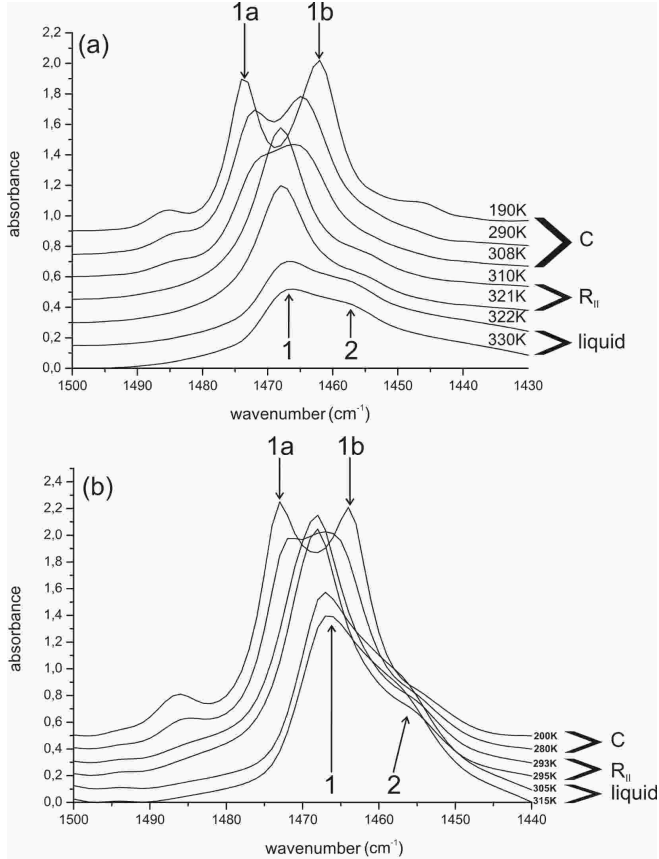}
 \caption{(a) IR spectrum showing the $\text{CH}_2$ scissor-vibration for bulk C$_{16}$H$_{33}$OH at
 various temperatures.
 (b) IR spectrum showing the $\text{CH}_2$ scissor-vibration of C$_{16}$H$_{33}$OH confined in mesoporous Si at
 various temperatures.
 \label{fig:CHspectrum}}
\end{figure}

\begin{figure}[ht]
\center\includegraphics[scale=0.5]{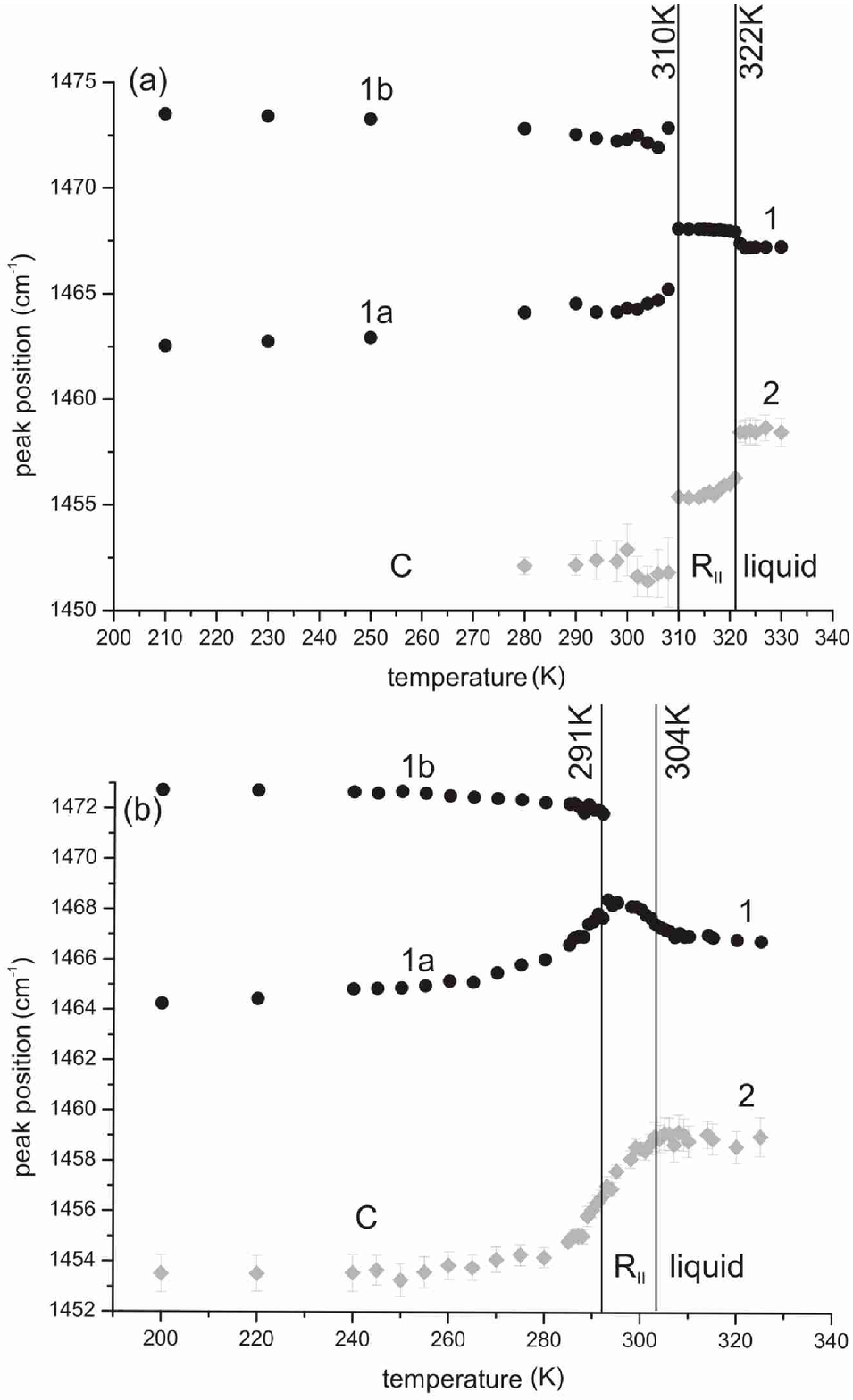}
 \caption{Wavenumber of the CH - scissor peak vs temperature for (a) bulk C$_{16}$H$_{33}$OH
 and (b) confined C$_{16}$H$_{33}$OH (the peak labels refer to Fig.~\ref{fig:CHspectrum} ).
 \label{fig:CHTemp} }
\end{figure}

\begin{figure}[ht]
\center\includegraphics[scale=0.485]{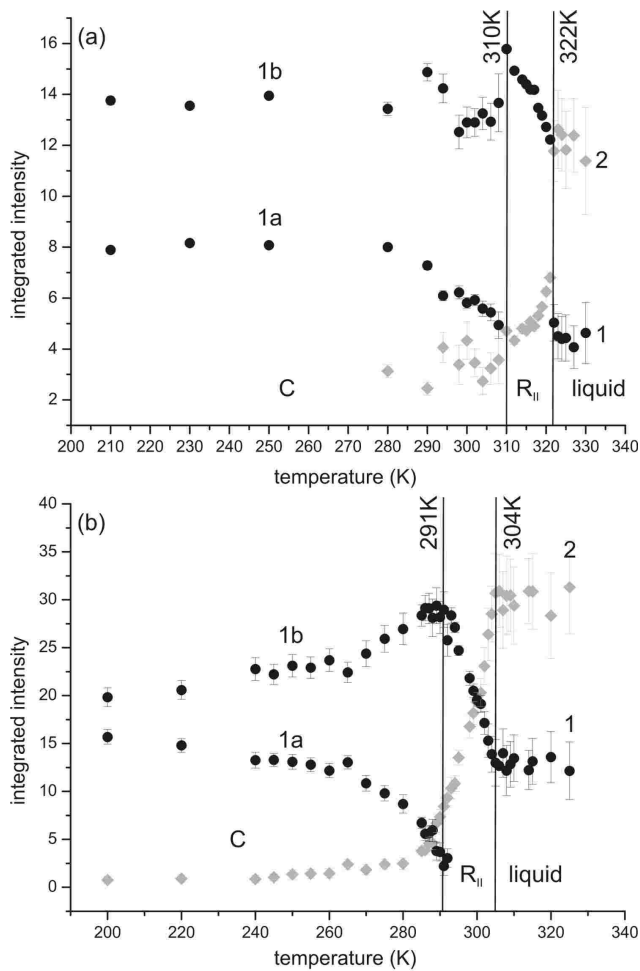}
 \caption{Integrated intensity of the CH - scissor peak vs temperature for (a) bulk C$_{16}$H$_{33}$OH
 and (b) confined C$_{16}$H$_{33}$OH (the peak labels refer to Fig.~\ref{fig:CHspectrum} ).
 \label{fig:CHInt} }
\end{figure}

Now let us turn towards the scissor-vibration of the $\text{CH}_2$
groups  (bending mode) that will give us information about
inner-molecular and inter-molecular force constants. The spectra
are shown in Fig.~\ref{fig:CHspectrum}. At first, we want to
discuss the bulk material. At high temperatures (liquid state) a
superposition of two peaks at 1467~cm$^{-1}$ and 1460~cm$^{-1}$ is
observed. In the intermediate temperature range ($R_{II}$ phase;
see Fig.~\ref{fig:bulkstructure_rot}) the intensity of the peak
labeld "1" increases strongly. At low temperatures (C phase; see
Fig.~\ref{fig:bulkstructure_cryst}) this band splits up into two
peaks. The latter transition can be clearly seen in
Figs.~\ref{fig:CHTemp}a) and \ref{fig:CHInt}a), where we display
the peak positions and intensities as a function of temperature.
The results are similar to those obtained for the bulk state of
n-paraffines, that apart from the missing OH-group are similar in
their structure, i.~e.\ that have the same
CH$_2$-backbone~\cite{Snyder1961}. In IR-spectra only one
CH$_2$-scissoring-band is observed at high temperatures, i.~e.\
intra-molecular interactions of the CH$_2$-groups are too small to
lead to a series of distinct peaks. The band splitting at low
temperatures has been attributed to inter-molecular interactions
(see Ref.~\cite{Snyder1961} and text below).

Qualitatively, a behavior similar to that of the bulk state  is
observed for confined C$_{16}$H$_{33}$OH (see
Fig.~\ref{fig:CHspectrum}b). In the high-temperature liquid phase two overlapping
peaks are visible. The stronger one, i.~e.\ that at higher wavenumbers, undergoes an
increase in intensity at about 304~K (see Fig.~\ref{fig:CHInt}b), indicating the transition
from the liquid phase to the $R_{II}$ phase, while the secondary peak at lower
wavenumbers gets weaker and finally
 disappears. At the second transition temperature of $T=291$~K
  the remaining strong peak splits (see also Fig.~\ref{fig:CHTemp}b).
  The separation is not as distinct as for bulk material. These transition
  temperatures, $T=304$~K and $T=291$~K (see Figs.~\ref{fig:CHTemp}b and
  \ref{fig:CHInt}b), agree well with those obtained via x-ray measurements
  (compare with Table~\ref{tab:MeltingPoints}).\\

\begin{table}[ht]
\begin{center}
\begin{tabular}{|c||c|c|c|c|}\hline
    & \multicolumn{2}{|c|}{bulk} & \multicolumn{2}{|c|}{confinement}\\\cline{2-5}
    & liquid & $R_{II}$ & liquid & $R_{II}$\\\hline \hline
  scissor [cm$^{-1}$]& 1467 & 1467 & 1467 & 1467\\\hline
  sym. stretch [cm$^{-1}$]& 2854 & 2851 & 2854 & 2851\\\hline
  assym. stretch [cm$^{-1}$]& 2927 & 2921 & 2924 & 2918\\\hline
  $f_d$ [N/m]& 455 & 453 & 454 & 452\\\hline
  $f_\alpha$ [N/m]& 56 $\pm$ 1 & 56 $\pm$ 1 & 57 $\pm$ 1 & 57 $\pm$ 1\\\hline
\end{tabular}
\caption{\label{tab:ergebnis1} Wavenumbers
$\overline{\nu}=\omega/(2\pi c)$ (with $\omega$ being the angular
frequency and $c$ the speed of light) and resulting stretching and
bending force constants in the liquid and $R_{II}$ phase of bulk
and confined C$_{16}$H$_{33}$OH. $f_\alpha$ has been evaluated
using both the Eq.~(\ref{eq:alpha1}) and the Eq.~(\ref{eq:alpha2}).
The difference yields the specified uncertainty. $f_\alpha$ are in
units of $N/m$ (see Eq.~(\ref{eq:pot2}) in Appendix~A and
Ref.~\cite{Meister1946}). To get $f_\alpha$ in units $Nm/rad^{2}$
one has to multiply $f_\alpha$ with d$^2$, where $d=1,09 \cdot
10^{-10}$~m is the CH bond length.}
\end{center}
\end{table}

\begin{figure}[ht]
 \center\includegraphics[scale=0.35]{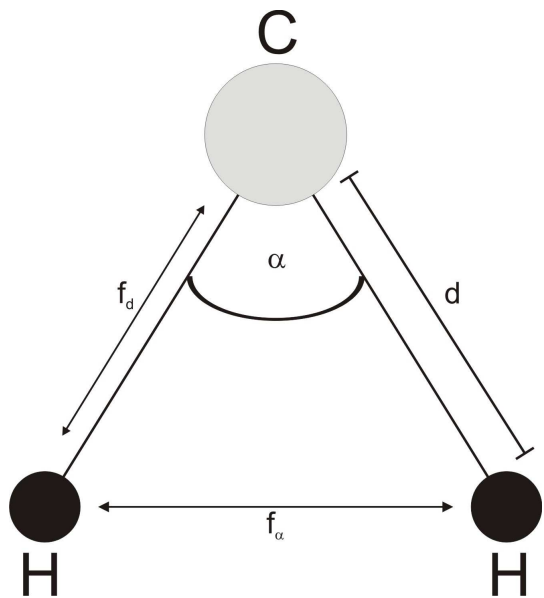}
 \caption{$\text{CH}_2$ molecules with C-H bondlength d, H-C-H angle $\alpha$ and the resulting
 inner force constants $f_d$ and $f_{\alpha}$\label{fig:moleculeCH2} }
\end{figure}

In the following we want to analyze the dynamics of the
CH$_2$-groups  {\bf in order to check whether it is affected by
the geometric confinement, e.~g.\ by an interaction with the pore
surfaces, by the limited number of neighboring molecules
(finite-size-effects) or by structural changes}. In a first
approximation we can assume that it is not affected by the
stretching of the OH - groups. On the one hand, there is the
scissor vibration, where the angle $\alpha$ between the two
CH-bonds oscillates around its equilibrium value $\alpha=109.47$°
(see Fig.~\ref{fig:moleculeCH2}). In addition, symmetric and
asymmetric stretching vibrations of the CH-bonds are observable
(for the values see Table~\ref{tab:ergebnis1}). Let $f_\alpha$ and
$f_d$ denote the respective force constants. These can be
calculated from the measured vibration frequencies using
Eqs.~(\ref{eq:fd})-(\ref{eq:alpha2}) (see Appendix A; the
difference in calculating $f_\alpha$ via Eq.~(\ref{eq:alpha1}) or
Eq.~(\ref{eq:alpha2}) is below 3.5\% confirming that the
inner-molecular coupling terms can be neglected).
Table~\ref{tab:ergebnis1} shows the results for the liquid and the
$R_{II}$ phase. Neither the phase transition liquid $\rightarrow$
R$_{II}$ nor geometrical confinement does markely affect the
innermolecular constants.

\begin{figure*}[ht]
 \center\includegraphics[scale=0.7]{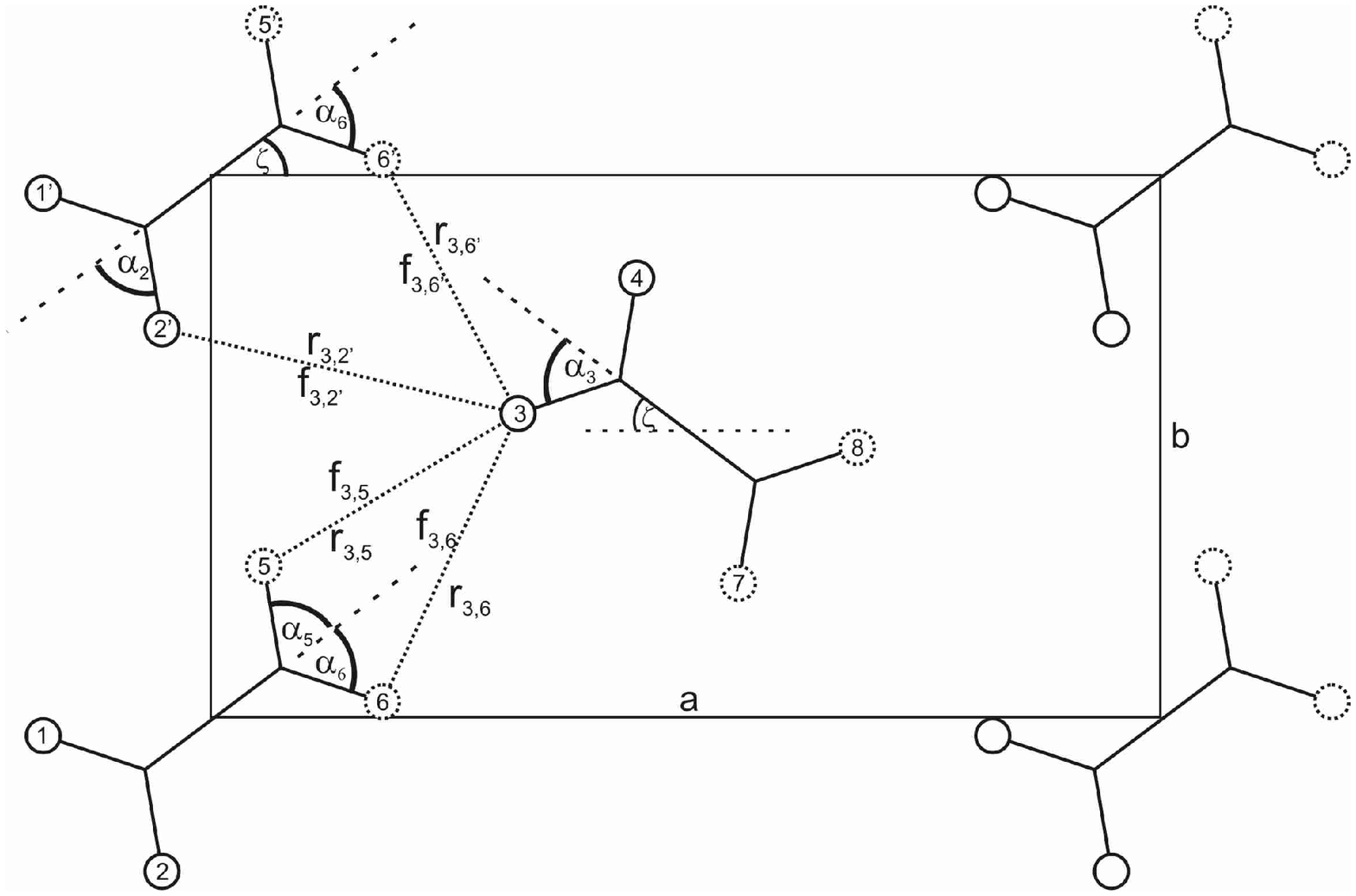}
 \caption{Lattice of C$_{16}$H$_{33}$OH in the
 C phase of the $\beta$-form, i.~e.\ a view on the a-b-plane
 perpendicular to the molecules axis (compare with
 Fig.~\ref{fig:confstructure_cryst}).
 Solid circles are
 H-atoms in the same plane, dashed circles are H-atoms lying in a plane
 above.
 \label{fig:lattice}}
\end{figure*}

In the $R_{II}$ phase the molecules rotate about their  long axis,
so that the primitive cell consists of only one molecule per layer
(see Fig.~\ref{fig:bulkstructure_rot}). Therefore, no splitting is
observed. But in the C phase (below 310~K for bulk and below 291~K
for confined C$_{16}$H$_{33}$OH), where the molecules are arranged
in a herringbone structure, there are two molecules per layer in
the primitive cell. So the symmetry of the arrangement allows a
splitting of the scissoring band and obviously the molecular
interactions are sufficiently strong that we are able to observe a
double peak (see above, Figs.~\ref{fig:CHspectrum} and
\ref{fig:CHTemp}). The strength of interaction depends on the
distances between neighboring H-atoms of adjacent chains and can
be analyzed using a formalism developed by Snyder (see
Ref.~\cite{Snyder1961} and Appendix B). In Fig.~\ref{fig:lattice}
we have sketched the orthorhombic lattice of the crystalline
C$_{16}$H$_{33}$OH subcell (a view on the a-b-plane perpendicular
to the molecules axis). In what follows we restrict ourselves to
this $\beta$-form, that is characteristic for confined
C$_{16}$H$_{33}$OH (a quantitative analysis of bulk
C$_{16}$H$_{33}$OH is difficult due to the superposition of
$\beta$- and $\gamma$-form). Assuming that the inner force
constant $f_\alpha$ does not change at the phase transition, the
intermolecular force constants $f_{3,j}$ can be evaluated from the
observed splitting of the scissor band as described in Appendix B
(see Eq.~(\ref{eq:fij})). The values needed are the lattice
parameters (see Table~\ref{tab:ab}) and the herringbone angle
$\zeta$ between the projection of the backbone and the $a$-axis
(see Fig.~\ref{fig:lattice}). The latter one is determined via
Eq.~(\ref{eq:Winkel}) and the measured intensities of the two
CH$_2$-scissoring-peaks. For confined C$_{16}$H$_{33}$OH we have
$I_a = 13.29$ and $I_b = 22.23$ yielding an angle of $\zeta =
37.7$° (see Fig.~\ref{fig:CHInt}b for $T=245$~K). We display the
intermolecular force constants in Table~\ref{tab:ergebnis4}. For
comparison, we also list literature values for an alkane,
C$_{23}$H$_{48}$ at 90~K, which have been evaluated in the same
way \cite{Snyder1961}. This alkane and C$_{16}$H$_{33}$OH exhibit
a similar structure: The backbones of the molecules consist of the
same CH$_2$-units and both take the $\beta$-form at
low-temperatures. In addition, also the values of the lattice
constants for C$_{23}$H$_{48}$, $a=7.45$~\r A and $b=4.96$~\r A,
are close to those of C$_{16}$H$_{33}$OH (see Tab.~\ref{tab:ab}).
Due to this structural similarity the intermolecular distances
listed in Table~\ref{tab:ergebnis4} are similar, however, the
respective force constants differ slightly by 10 to 20\%. This is
mainly due to the orientation of CH$_2$-groups (the projection of
the backbones on the a-b-plane) characterized by the herringbone
angle $\zeta$. For C$_{16}$H$_{33}$OH $\zeta=37.7$° holds, for the
alkane $\zeta=42$°.

This difference is probably due to the presence of polar OH-groups
in C$_{16}$H$_{33}$OH that are strongly interacting and thus have an
impact on the molecular orientation. The above comparison confirms
once again that confined C$_{16}$H$_{33}$OH takes the $\beta$-form in
contrast to the bulk material ($\gamma$- and $\beta$-form).

In order to assess the validity of our analysis, we also calculate
the theoretical band splitting of the CH$_2$-scissoring vibration
and compare it with the measured values. Using the values from
Table \ref{tab:ergebnis4} as well as Eqs.~(\ref{eq:splitting}) and
(\ref{eq:Gab}), we get a theoretical value of $\Delta\overline{\nu}_{calc}=8.1$~cm$^{-1}$
for confined C$_{16}$H$_{33}$OH at $T = 245$ K. The measured band splitting is
$\Delta\overline{\nu}_{meas}=7.8$~cm$^{-1}$. Therefore, the
experimental data is in good agreement with the theory.

\begin{table}[ht]
\begin{center}
\footnotesize{
\begin{tabular}{|c|c|c|c||c|c|}\hline
  H-Atoms & force    & \multicolumn{2}{|c||}{confined C$_{16}$H$_{33}$OH} & \multicolumn{2}{|c|}{bulk C$_{23}$H$_{48}$}\\
          & constant & \multicolumn{2}{|c||}{(this work)} &
          \multicolumn{2}{|c|}{(Ref.~\cite{Snyder1961})}\\
       &      & \multicolumn{2}{|c||}{$\zeta=37.7$°} &
       \multicolumn{2}{|c|}{$\zeta=42$°}          \\\hline
      &       & \scriptsize{distances (\AA)} & \scriptsize{$10^{-21}$~Nm} &
      \scriptsize{distances (\AA)} & \scriptsize{$10^{-21}$~Nm}\\\hline
  3-2'& $f_{3,2'}$ & 2.85 & -4.674 & 2.79 & -5.837\\\hline
  3-5 & $f_{3,5}$ & 2.59 & -2.775 & 2.70 & -2.376\\\hline
  3-6 & $f_{3,6}$ & 2.96 & -3.508 & 2.94 & -3.920\\\hline
  3-6'& $f_{3,6'}$ & 2.96 & -3.508 & 2.94 & -3.920\\\hline
\end{tabular}
}
\end{center}
\caption{\label{tab:ergebnis4} Distances between two neighboring
hydrogen atoms and the resulting interaction force constants in
the C phase confined C$_{16}$H$_{33}$OH ($\beta$-form). $\zeta$
denotes the herringbone angle between the projection of the backbone and the
a-axis (see Fig.~\ref{fig:lattice}). For comparison, we also list
literature values for an alkane, C$_{23}$H$_{48}$ at 90~K, that
also takes the $\beta$-form \cite{Snyder1961}. The CH$_2$-backbones
of both molecules are similar, but C$_{16}$H$_{33}$OH exhibits an
additional polar OH-group that affects the orientation angle
$\zeta$.}
\end{table}

\section{Summary}
We have studied the structure and molecular dynamics of
n-hexadecanol  confined in nanochannels of mesoporous silicon and
of bulk n-hexadecanol in their respective phases (in the order of
decreasing temperature: liquid, rotator R$_{II}$ and C). For this
purpose we have performed x-ray and infrared-measurements.

The transition-temperatures for confined C$_{16}$H$_{33}$OH are
lower  than for bulk C$_{16}$H$_{33}$OH ($\Delta T \simeq 20$~K,
see Table~\ref{tab:MeltingPoints}). In addition, under confinement
the phase transitions are smeared, probably due to a distribution
of pore diameters. Geometrical confinement does not affect the
innermolecular force constants of the CH$_2$-scissoring vibration
(see Table~\ref{tab:ergebnis1}) but has an impact on the molecular
arrangement. The R$_{II}$ phase of both bulk and confined
hexadecanol is characterized by an orthorhombic subcell, where the
chain axis are parallel to the layer normal (see
Fig.~\ref{fig:bulkstructure_rot}).  However, in the
low-temperature C phase there is a fundamental structural
difference. While bulk C$_{16}$H$_{33}$OH  exhibits  a
polycrystalline mixture of $\beta$- and $\gamma$-forms (see
Figs.~\ref{fig:bulkstructure_cryst} and
\ref{fig:confstructure_cryst}), geometrical confinement favors a
phase closely related to the $\beta$-form: only crystallites with
an orthorhombic subcell are formed, where the chain axes are
parallel to the bi-layer normal. However, the $\gamma$-form having
a monoclinic subcell, in which the chain axis are tilted with
respect to the layer normal, is suppressed. A reason for this
might be the irregular shape of the nanochannels, into which the
crystallites have to fit, favoring the formation of the
geometrically more simple and less bulky form
\cite{Christenson2001,Morishige2000}~(see
Fig.~\ref{Porenschnitt}). Since only the pure $\beta$-form is
present under confinement, we were able to evaluate the
inter-molecular force constants of the CH$_2$-scissor vibration.
Also the orientation of the $\beta$-crystallites has been
determined: the molecules are arranged with their long axis
perpendicular to the pore axis.

\newpage\noindent
\section*{Appendix A}
In this section we show how the innermolecular force constants of
the CH$_2$-groups can be evaluated using three characteristic
vibration frequencies, that are easily measured: the scissor
vibration as well as the symmetric and asymmetric CH-bond
stretching. For this purpose we apply the Wilson FG - matrix
method~\cite{Wilson1939}. We use the notation of Meister and
Cleveland for the similar $\text{H}_2\text{O}$ molecule
\cite{Meister1946}
and perform the calculations in the same way.\\

Fig.~\ref{fig:moleculeCH2} shows a single $\text{CH}_2$ -
molecule. In the following we will neglect the influence of the
neighboring molecules on this one. $d=1.09$~\r A is the
length of the C - H bond and $\alpha = 109.47$° the angle between
the two C - H bonds \cite{Abrahamsson1960}.
 This kind of molecule
belongs to the $C_{2_{\nu}}$ point group. This means, there are
two vibrations of type $A_1$ (symmetric stretching and bending
vibration) and one vibration of type $B_2$ (asymmetric
stretching vibration). The internal coordinates of this molecule
are $\Delta d_1$, $\Delta d_2$ and $\Delta\alpha$. $\Delta d_1$
and $\Delta d_2$ mean changes in the bond length of the two C - H
- bonds and $\Delta\alpha$ changes in the angle between the two
bonds. Therefore we get three symmetry coordinates, two for $A_1$
and one for $B_2$. If we assume d being the equilibrium
C - H bond length, then we obtain for the three symmetry coordinates:
\begin{equation}
R_1=\sqrt{\frac{1}{2}}\Delta d_1+\sqrt{\frac{1}{2}}\Delta d_2\\
\end{equation}
\begin{equation}
R_2=\Delta\alpha \cdot d
\end{equation}
\begin{equation}
R_3=\sqrt{\frac{1}{2}}\Delta d_1-\sqrt{\frac{1}{2}}\Delta d_2
\end{equation}\\
Now, we have to calculate the \textbf{F} matrix, related to the
potential energy, and the \textbf{G} matrix related to the kinetic
energy. The potential energy can be written as
\begin{equation}
2V=\sum f_{ik}r_ir_k
\label{eq:pot1}
\end{equation}
and with the internal coordinates
\begin{equation}
\begin{split}
2V=&f_d\left[\left(\Delta d_1\right)^2+\left(\Delta d_2\right)^2\right]\\
&+f_{\alpha}(d\Delta\alpha)^2+2f_{d\alpha}\left(\Delta d_1+\Delta d_2\right)\left(d\Delta\alpha\right)\\
&+2f_{dd}\left(\Delta d_1\right)\left(\Delta d_2\right)
\label{eq:pot2}
\end{split}
\end{equation}
Now we set $d_1=d_2=d$ and write Eq.~(\ref{eq:pot2}) as
\begin{equation}
2V=\sum F_{jl}R_jR_l
\end{equation}
with $F_{jl}=F_{lj}$.
In matrix form, Eqs.~(\ref{eq:pot1}) and (\ref{eq:pot2}) become
\begin{equation}
2V = \textbf{r}'\textbf{fr}
\label{eq:vmatrix1}
\end{equation}
and
\begin{equation}
2V=\textbf{R}'\textbf{FR}
\label{eq:vmatrix2}
\end{equation}
\textbf{r}$'$ and \textbf{R}$'$ are the transposes of \textbf{r} and \textbf{R}. With Eqs.~(\ref{eq:vmatrix1}) + (\ref{eq:vmatrix2})
\begin{equation}
\textbf{r}'\textbf{fr}=\textbf{R}'\textbf{FR}
\end{equation}
The $R_i$'s are linear combinations of the $r_i$'s
\begin{equation}
\begin{split}
R_i&=\sum_{k}U_{ik}r_k\\
\textbf{R}&=\textbf{Ur}
\end{split}
\end{equation}
Since the $R_i$'s are orthogonal and normalized, then
$\textbf{U}^{-1}=\textbf{U}'$ and
\begin{eqnarray}
\textbf{r}=\textbf{U}'\textbf{R}\\
\textbf{r}'=(\textbf{U}'\textbf{R})'=\textbf{R}'\textbf{U}
\end{eqnarray}
This means with Eqn. (10)
\begin{eqnarray}
\textbf{R}'(\textbf{UfU}')\textbf{R}=\textbf{R}'\textbf{FR}\\
\textbf{F}=\textbf{UfU}'
\end{eqnarray}
The \textbf{F} matrix is
\begin{center}
\begin{math}
\begin{array}{c|ccc}
 & \Delta d_1 & \Delta d_2 & \Delta\alpha\\\hline
 \Delta d_1 & f_d & f_{dd} & df_{d\alpha}\\
 \Delta d_2 & f_{dd} & f_d & df_{d\alpha}\\
 \Delta\alpha & df_{d\alpha} & df_{d\alpha} & d^2f_{\alpha}
\end{array}
\end{math}
\end{center}
The \textbf{U} matrix for type $A_1$ is\\
\begin{center}
\begin{math}
\begin{array}{c|ccc}
A_1 & \Delta d_1 & \Delta d_2 & \Delta\alpha\\\hline
R_1 & \sqrt{\frac{1}{2}} & \sqrt{\frac{1}{2}} & 0\\
R_2 & 0 & 0 & 1
\end{array}
\end{math}
\end{center}
and for $B_2$
\begin{center}
\begin{math}
\begin{array}{c|ccc}
B_2 & \Delta d_1 & \Delta d_2 & \Delta\alpha\\\hline
R_3 & \sqrt{\frac{1}{2}} & -\sqrt{\frac{1}{2}} & 0
\end{array}
\end{math}
\end{center}
So, for the type $A_1$ the \textbf{F} matrix is\\
\begin{equation}
\textbf{F}_{A_1} = \textbf{UfU}'=
\left(\begin{array}{cc}
F_{11} & F_{12}\\
F_{21} & F_{22}
\end{array}\right)
=
\left(
\begin{array}{cc}
f_d + f_{dd} & \sqrt{2}df_{\alpha}\\
\sqrt{2}df_{\alpha} & d^2f_{\alpha}
\end{array}\right)
\end{equation}
and for the $B_2$ type
\begin{equation}
\textbf{F}_{B_2}=(F_{33})=(f_d-f_{dd})
\end{equation}
The exact derivation of the $\textbf{G}$ matrix shouldn't be shown here.
It can be gleaned by Meister and Cleveland \cite{Meister1946}.
Only the most important steps shall be explained here.\\
If only non-degenerate vibrations are present, the elements of the
kinetic energy matrix can be written as
\begin{equation}
G_{jl}=\sum_{p}\mu_pg_p\textbf{S}_j^{(t)}\textbf{S}_l^{(t)}
\end{equation}
where j and l refer to symmetry coordinates used in determining
the \textbf{S} vector, $p$ refer to a set of equivalent atoms, a
typical one of the set being t. $\mu_p$ is the reciprocal of the
mass of the typical atom $t_p$ and $g_p$ is the number of
equivalent atoms in the $p$th set. The \textbf{S} vector is given by
\begin{equation}
\textbf{S}_j^{(t)}=\sum_{k}U_{jk}s_{kt}
\end{equation}
where $j$, $U_jk$ and $\sum_k$ have the same meaning as above. $s_{kt}$
can be expressed in terms of unit vectors along the chemical bonds and depends
on the changes in the bond length or the angle between the bonds.
So, the \textbf{G} matrix for the $A_1$ vibration type has the form
\begin{equation}
\begin{split}
\textbf{G}_{A_1}&=
\left(\begin{array}{cc}
G_{11} & G_{12}\\
G_{21} & G_{22}
\end{array}\right)\\
&=\left(\begin{array}{cc}
\mu_H+\mu_C(1+\cos\alpha) & -\frac{\mu_C\sqrt(2)\sin\alpha}{d}\\
-\frac{\mu_C\sqrt(2)\sin\alpha}{d} & \frac{2\mu_H+\mu_C(1-\cos\alpha)}{d}
\end{array}\right)
\end{split}
\end{equation}
and for the $B_2$ vibration type
\begin{equation}
\textbf{G}_{B_2}
(G_{33})=(\mu_H+\mu_C(1-\cos\alpha))
\end{equation}
To determine the frequencies, one has to solve the equation
\begin{equation}
\left|\textbf{GF}-\lambda\textbf{E}\right|=0 \qquad
\end{equation}
where $\lambda = \omega^2= (\overline{\nu} 2 \pi c)^2$ denotes the square of the angular frequency. For the $A_1$ type one gets
the equation
\begin{equation}
\begin{split}
\lambda^2&-\lambda(F_{11}G_{11}+2F_{12}G_{12}+F_{22}G_{22})\\
&+\begin{array}{|cc|}
F_{11}&F_{12}\\
F_{21}&F_{22}
\end{array}
\cdot
\begin{array}{|cc|}
G_{11}&G_{12}\\
G_{21}&G_{22}
\end{array}=0
\label{eq:loesung1}
\end{split}
\end{equation}
and for the $B_2$ type
\begin{equation}
\lambda_3-F_{33}G_{33}=0
\label{eq:loesung2}
\end{equation}
Eq.~(\ref{eq:loesung1}) can be separated with the Vieta
expression \cite{Siebert1966}. Inserting the terms for the $F_{ij}$
and $G_{ij}$, we obtain
\begin{eqnarray}
\begin{aligned}
\lambda_1+\lambda_2 & = (f_d+f_{dd})[\mu_C(1+\cos\alpha)+\mu_H]\\
&
+2f_{\alpha}[\mu_C(1-\cos\alpha)+\mu_H]-4f_{d\alpha}\mu_C\sin\alpha
\label{l1+l2}
\\
\lambda_1\cdot\lambda_2 &=
[(f_d+f_{dd})f_{\alpha}-2f_{d\alpha}^2]2\mu_H(2\mu_C+\mu_H)
\end{aligned}
\label{lamb1lamb2}
\end{eqnarray}
For Eq.~(\ref{eq:loesung2}) one obtains
\begin{equation}
\lambda_3=(f_d-f_{dd})[\mu_C(1-\cos\alpha)+\mu_H] \label{lamb3}
\end{equation}

Neglecting the coupling constants $f_{dd}$ and $f_{d\alpha}$
allows to evaluate the innermolecular force constants using the
measured wave numbers, $\overline{\nu}_{d,sym}=\sqrt{\lambda_1}/(2\pi c)$,
$\overline{\nu}_\alpha=\sqrt{\lambda_2}/(2\pi c)$ and
$\overline{\nu}_{d,asym}=\sqrt{\lambda_3}/(2\pi c)$. Then Eq.~(\ref{lamb3}) yields

\begin{equation}
f_d= (2\pi c)^2 \cdot \frac{\overline{\nu}_{d,
asym}^2}{\mu_C(1-\cos\alpha)+\mu_H}\label{eq:fd}
\end{equation}
Inserting this result into Eq.~(\ref{lamb1lamb2}) yields
\begin{equation}
\label{eq:alpha1} f_{\alpha}= (2\pi c)^2 \cdot
\frac{\overline{\nu}_{d,sym}^2\cdot \overline{\nu}_\alpha^2}
{\overline{\nu}_{d,asym}^2} \cdot
\frac{\mu_C(1-\cos\alpha)+\mu_H}{2\mu_H(2\mu_C+\mu_H)}
\end{equation}
There is a second possibility to evaluate $f_\alpha$, i.~e.\ by
inserting Eq.~(\ref{eq:fd}) into Eq.~(\ref{l1+l2}). This yields
\begin{equation}
f_{\alpha}= (2\pi c)^2 \cdot \frac{\overline{\nu}_{d,sym}^2 +
\overline{\nu}_\alpha^2- \overline{\nu}_{d,asym}^2 \cdot
\frac{\mu_C(1+\cos\alpha)+\mu_H}{\mu_C(1-\cos\alpha)+\mu_H}}{2(\mu_C(1-\cos\alpha)+\mu_H)}
\label{eq:alpha2}
\end{equation}

Taking the measured wavenumbers listed in
Table~\ref{tab:ergebnis1} and the average angle between the
CH-bonds, $\alpha=109.4$°, as well as the masses of the atoms,
$1/\mu_C=12u$ and $1/\mu_H=1u$ ($u=1.6606 \cdot 10^{-27}$~kg),
Eqs.~(\ref{eq:fd})-(\ref{eq:alpha2}) yield the force constants listed
in Table~\ref{tab:ergebnis1}. The difference in calculating
$f_\alpha$ via Eq.~(\ref{eq:alpha1}) or Eq.~(\ref{eq:alpha2}) is below
3.1\% confirming that the inner-molecular coupling terms can be
neglected.

\section*{Appendix B}


What follows is a summary of Snyder's derivation of the
intermolecular force constants between the CH$_2$ groups of
neighboring molecules that gives rise to a splitting of the
scissor band at low temperatures \cite{Snyder1961}. We show how
this formalism can be applied to C$_{16}$H$_{33}$OH. An
alternative description can be found in Ref.~\cite{Tasumi1965}.

In Fig.~\ref{fig:lattice} we display a hexagonal subcell of
C$_{16}$H$_{33}$OH. While Stein \cite{Stein1955} has taken only one
pair of neighboring $\text{CH}_2$ into account to calculate the
splitting of rocking and scissoring bands, Snyder has shown that
more pairs have to be included. When we consider the distances of
H-atoms from the H-atom no.~3 (see Fig.~\ref{fig:lattice}),
then all atoms except no.~$2'$, $6'$, 5 and 6 have distances
larger then 3.7~\AA. The internal coordinates $\alpha_i$ are
always half of the angle between the C - H bonds of a
$\text{CH}_2$ molecule. Solid circles are H-atoms in the same
plane, dashed circles H-atoms in a plane above or below.
\begin{figure}[ht]
 \center\includegraphics[scale=0.35]{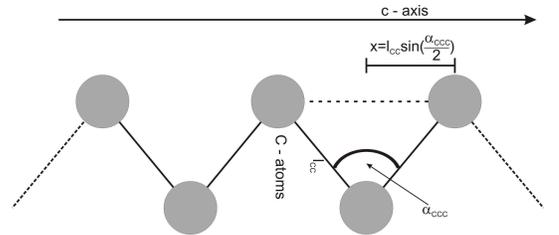}
 \caption{Lateral view at the long axis of the C$_{16}$H$_{33}$OH chain,
 x is the projection of the C - C distance on the c-axis of the
 crystal lattice
 \label{fig:Kette}}
\end{figure}\\
Now, we want to write the positions of these five H-atoms as a
vector. Fig.~\ref{fig:Kette} shows the lateral view of a part of
the C$_{16}$H$_{33}$OH chain. With values from Abrahamsson
\cite{Abrahamsson1960} for $l_{CC}=1.545$~\r A and
$\alpha_{CCC}=110.4$°, we can calculate the distance of the
a-b-plane to the corresponding plane above or below with
\begin{equation}\nonumber
x=l_{CC}\sin(\frac{\alpha_{CCC}}{2})=1.2687\,\text{\AA}
\end{equation}
Assuming that the hydrogen in the central plane has the $c$
component 0, the hydrogen in the plane above has the component
$c=1.2687$~\AA.
\begin{figure}[ht]
 \center\includegraphics[scale=0.3]{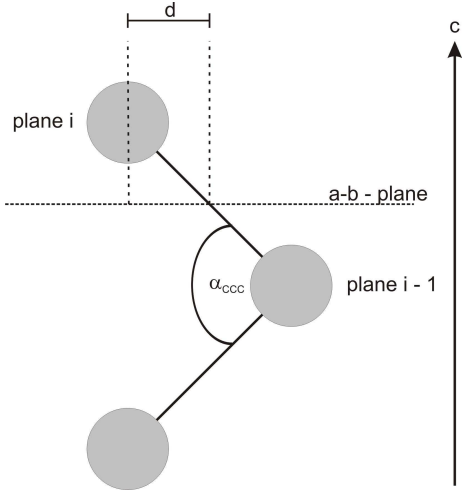}
 \caption{Lateral view at the long axis of the C$_{16}$H$_{33}$OH chain,
 d is the projection of half of the C - C distance on the a-b-plane of the
 subcell.
 \label{fig:seitlich}}
\end{figure}\\
The projection of the C - C bond in the a-b-plane is according to
Fig.~\ref{fig:seitlich}
\begin{equation}\nonumber
d=\frac{l_{CC}}{2}\cos(\frac{\alpha_{CCC}}{2})=0.4409\,\text{\AA}
\end{equation}
Taking the point \textbf{0} for the lower left edge of the ab -
plane the five atoms have the coordinates:
\begin{equation}\nonumber
\begin{split}
H_3&=
\left(\begin{array}{c}
\frac{a}{2}-d\cos(\zeta)-l\cos(\alpha_3-\zeta)\\
\frac{b}{2}+d\sin(\zeta)-l\sin(\alpha_3-\zeta)\\
0
\end{array}\right)\\
\nonumber
H_{2'}&=
\left(\begin{array}{c}
-d\cos(\zeta)+l\sin(\alpha_{2'}-\frac{\pi}{2}+\zeta)\\
b-d\sin(\zeta)-l\cos(\alpha_{2'}-\frac{\pi}{2}+\zeta)\\
0
\end{array}\right)\\
\nonumber
H_{5}&=
\left(\begin{array}{c}
d\cos(\zeta)-l\sin(\alpha_{5}-\frac{\pi}{2}+\zeta)\\
d\sin(\zeta)+l\cos(\alpha_{5}-\frac{\pi}{2}+\zeta)\\
1.2674
\end{array}\right)\\
\nonumber
H_{6}&=
\left(\begin{array}{c}
d\cos(\zeta)+l\cos(\alpha_6-\zeta)\\
d\sin(\zeta)-l\sin(\alpha_6-\zeta)\\
1.2674
\end{array}\right)\\
\nonumber
H_{6'}&=
\left(\begin{array}{c}
d\cos(\zeta)+l\cos(\alpha_{6'}-\zeta)\\
b+d\sin(\zeta)-l\cos(\alpha_{6'}-\zeta)\\
1.2674
\end{array}\right)
\end{split}
\end{equation}
with $a$ and $b$ being the lattice constants of the crystalline
phase. We get the distances between the atom 3 and the other ones
(see Fig.~\ref{fig:lattice}) with
\begin{equation}
r_{3j}=\left|H_3-H_j\right| \label{eq:distances}
\end{equation}
where $j=2', 6', 5, 6$ is. The herringbone angle $\zeta$ (see the
upper left corner of Fig.~\ref{fig:lattice}) can be determined
with the relation \cite{Snyder1961}
\begin{equation}
\frac{I_a}{I_b}=\tan^2\zeta
\label{eq:Winkel}
\end{equation}
where $I_a$ is the integrated intensity of the scissoring mode, which is polarized in
the a direction (higher mode at 1473~cm$^{-1}$) and $I_b$ the one of the mode,
which is polarized in the $b$ direction (lower mode at 1462~cm$^{-1}$).\\

With the distances of two hydrogen atoms $H_3$ and $H_j$
(=$r_{3,j}$) [in our case 3 denotes the central H-atom (see
Fig.~\ref{fig:lattice}) and $j=2',5,6,6'$ the neighboring H-atoms
that interact] we obtain the intermolecular force constants
$f_{3,j}$:

\begin{eqnarray}
f_{3,j} & = &
\frac{\partial^2
V_{HH}}{\partial\alpha_3\partial\alpha_j} \nonumber \\
& = &
\left(\frac{\partial^2 V_{HH}}{\partial
r^2}\right)_{r_{3j}}\left(\frac{\partial r}{\partial
\alpha_3}\right)\left(\frac{\partial r}{\partial \alpha_j}\right)
 \label{eq:fij}
\end{eqnarray}
where $V_{HH}$ is the hydrogen repulsion potential introduced by
Dows~\cite{Dows1960}:
\begin{equation}
V_{HH}=1.2\cdot 10^{-10}e^{-3.52r}
\label{eq:potential}
\end{equation}
with r in \AA.

The values of $(\partial^2
V_{HH}/\partial r^2)_{r_{ij}}$ are obtained from
Eq.~(\ref{eq:beta}).

\begin{equation}
\beta=\frac{\partial^2 V_{HH}}{\partial r^2}=1.486848\cdot
10^{-9}e^{-3.52r} \label{eq:beta}
\end{equation}
with $\beta$ in
$\frac{ergs}{\text{\AA}^2}=10^{16}\frac{dyne}{cm}=10^{13}\frac{N}{m}$.

The measured intensity ratio (Eq.~(\ref{eq:Winkel})) allows us to
calculate the distances $r_{ij}$ as well as the partial
derivatives $\partial r/\partial \alpha_i$ (see
Eq.~(\ref{eq:distances}) and above). Finally, by knowing the
intermolecular force constants $f_{3,j}$ from Eq.~(\ref{eq:beta}) we
can evaluate the band splitting of the scissoring
vibration~\cite{Snyder1961}. For the angular frequencies
\begin{equation}
\overline{\nu}_1^2 - \overline{\nu}_2^2 = \left( \frac{1}{2 \pi
c}\right)^2 \cdot \underbrace{G_a^B\cdot \left\{
2f_{3,2'}-2(f_{3,6}+f_{3,6'})-4f_{3,5} \right\}}_{\Delta\lambda^B}
\label{eq:splitting}
\end{equation}
holds \footnote{In Snyder's general theory the force constants for
the scissoring vibration are denominated as $f_a^3$ (=
$f_{3,2'}$), $f_b^2$ (= $f_{3,6}+f_{3,6'}$) and $f_b^3$ (=
$f_{3,5}$).} with

\begin{equation}
G_a^B =\frac{4}{3}Q_R^2\mu_C +Q_r^2\mu_H \qquad
\label{eq:Gab}
\end{equation}

Here $1/\mu_C=12u$ and $1/\mu_H=1u$ denote the masses of the atoms
($u=1.6606 \cdot 10^{-27}$~kg), $1/Q_R=1.545 \cdot 10^{-10}$~m the
 C-C distance and $1/Q_r=1.09 \cdot 10^{-10}$~m the C-H
distance, so that $G_a^B=(0.88825/u)$~\AA$^{-2}= 5.349 \cdot
10^{46}$~(N m)$^{-1}$s$^{-2}$. For the band splitting of the
wavenumbers we thus obtain

\begin{equation}
\begin{split}
\Delta \overline{\nu}&=\overline{\nu}_1 -
\overline{\nu}_2\\
&= \left( \frac{1}{2 \pi c}\right)^2 \cdot
\frac{ G_a^B \cdot \left\{ 2f_{3,2'}-2(f_{3,6}+f_{3,6'})-4f_{3,5}
\right\} }{ \overline{\nu}_1 + \overline{\nu}_2}
\end{split}
\end{equation}

Inserting the values of Table~\ref{tab:ergebnis4}, i.~e.\
%
%
$(2f_{3,2'}-2(f_{3,6}+f_{3,6'})-4f_{3,5})=15.784 \cdot
10^{-21}$~Nm for confined C$_{16}$H$_{33}$OH, as well as the
respective wave numbers, that we take from Fig.~\ref{fig:CHTemp}
at low temperatures (labels 1a and 1b: $\overline{\nu}_1 +
\overline{\nu}_2 \simeq 2 \cdot 146700$~1/m) we get
$\Delta\overline{\nu}=810$~m
$^{-1}\equiv 8.1$~cm$^{-1}$ for confined C$_{16}$H$_{33}$OH.

\end{document}